\documentclass[conference]{IEEEtran}

\ifCLASSINFOpdf
\else
\fi
\usepackage{graphicx}
\usepackage{setspace}
\usepackage[cmex10]{amsmath}
\usepackage{cite}
\usepackage{url}
\usepackage{amsthm}
\usepackage{amssymb}
\usepackage{bbm}
\usepackage{eufrak}

\newtheorem{definition}{Definition}
\usepackage{algorithm}
\usepackage{algpseudocode}
\DeclareMathOperator*{\argmin}{arg\;min}

\newcommand{\matr}[1]{\mathbf{#1}} 

\usepackage{epstopdf}

\begin{document}

\title{Energy-Aware Relay Selection and Power Allocation for  Multiple-User Cooperative Networks}
\author{\IEEEauthorblockN{ Sabyasachi Gupta\IEEEauthorrefmark{1}, Ranjan Bose\IEEEauthorrefmark{1}\IEEEauthorrefmark{2}}
\IEEEauthorblockA{\IEEEauthorrefmark{1} Bharti School of Telecommunications, Indian Institute of Technology Delhi, New Delhi, India}
\IEEEauthorblockA{\IEEEauthorrefmark{2} Department of Electrical Engineering, Indian Institute of Technology Delhi, New Delhi, India} Email: sabyasachi.gupta@dbst.iitd.ac.in,
rbose@ee.iitd.ac.in
}

\maketitle

\begin{abstract}

This paper investigates the relay assignment and power allocation problem for two different network power management policies: group lifetime maximization (GLM) and minimum weighted total power (MWTP), with the aim of lifetime maximization in symbol error rate (SER) constrained multiple-user cooperative network. With optimal power allocation solution obtained for each policy, we show that the optimal relay assignment can be obtained using bottleneck matching (BM) algorithm and minimum weighted matching (MWM)  algorithm for GLM and MWTP policies, respectively. Since relay assignment with BM algorithm is not power efficient, we propose a novel minimum bottleneck matching (MBM) algorithm to solve the relay assignment problem optimally for GLM policy. To further reduce the complexity of the relay assignment, we propose suboptimal relay selection (SRS) algorithm which has linear complexity in number of source and relay nodes. Simulation results demonstrate that the proposed relay  selection and power allocation  strategies based on GLM policy have better network lifetime performance over the strategies based on MWTP policy.  Compared to the MBM and SRS algorithm, relay assignment based on BM algorithm for GLM policy has inferior network lifetime performance at low update interval. We show that relay assignment based on MBM algorithm achieves maximum network lifetime performance and  relay assignment with SRS algorithm has performance very close to it.


\end{abstract}

\IEEEpeerreviewmaketitle

\section{Introduction} 
User cooperation is a promising approach to achieve spatial diversity in wireless networks, where multiple antennas at nodes of the network are not available \cite{coop_protocols_laneman}. With nodes helping each other to transmit message through multiple independent fading paths, user cooperation  reduces the probability of erroneous message reception significantly, thus reducing the transmit power consumption in energy limited wireless network. To further achieve power efficiency in cooperative networks, several sum power minimization based power allocation strategies for a quality of service (QoS) constraint have been investigated  \cite{coop_pa_1rly_pwr_constr_outage, coop_pa_1_rs_rate_constr}. In many wireless networks maximizing network lifetime is the main design objective. Minimizing overall power in the network does not necessarily maximize network lifetime, since lifetime depends upon power minimization as well as energy balancing. Studies on  lifetime maximization with relay selection and power allocation  in single user cooperative network is available in \cite{coop_pa_1_rs_rate_constr, coop_pa_1_rs_rate_constrained2, coop_pa_relay_select_outage_constrained2, prediction_lifetime_coop_rate, maham_ser_constr_coop_journal}. Recently lifetime maximization in multiple-user network is studied in \cite{lifetime_cooperative_rate_constrained, lifetime_coop_multiuser_rly_select}. The study in \cite{lifetime_cooperative_rate_constrained} consider relay power allocation problem for lifetime maximization in rate constrained network. Relay assignment problem in multiple-user network without power allocation optimization is considered in \cite{lifetime_coop_multiuser_rly_select}.

In this paper we study the relay assignment and power allocation optimization problem for two network power management policies: group lifetime maximization (GLM) and minimum weighted total power (MWTP) to maximize lifetime of symbol error rate (SER) constrained multiple-user cooperative network. The relay selection and power allocation according to these policies are solved in two steps. At first, according to a policy the optimal power allocation solution is derived and a weight is assigned for each source relay pair. Then using this weight, relay assignment problem is solved optimally and suboptimally. The main contribution of this paper are threefold. Firstly, we consider relay assignment along with source relay power allocation for lifetime maximization in multiple-user network which has not been studied before to the best of our knowledge. Secondly, we solve  relay assignment problem in multiple-user network optimally and suboptimally with lower complexity compared to the proposed method in \cite{lifetime_coop_multiuser_rly_select}. Thirdly we show that compared to the MWTP policy proposed in \cite{coop_pa_1_rs_rate_constrained2, lifetime_cooperative_rate_constrained} for relay selection and power allocation for lifetime maximization, GLM policy based strategies achieve higher network lifetime.

\section{System Model and Problem Formulation}
\label{sec_system model}
\subsection{System Model}
Consider a wireless network consisting of set of $M$ source nodes $S=\{s_1, .. s_M\}$ and $N$ relay nodes $R=\{r_1, ..,r_N \}$ randomly distributed in an area. Each node is equipped with a single omnidirectional antenna and a battery energy supply. Each source node sends data to the base station (BS), denoted by $d$, with the help of a relay node selected for the source node. Each relay can at most be selected by only one source node. Therefore, we need $N \geq M$. To avoid interference, each source data transmission is assigned orthogonal channels using frequency-division multiple access. The source-BS, source-relay and relay-BS channels undergo independent quasi-static Rayleigh fading and path loss. The channel variances of source node $s_i \in S$ to BS, source node $s_i$ to relay node $r_j \in R$ and relay node $r_j$ to BS are modelled as $\sigma_{s_i,d}^2=\eta D_{s_i,d}^{-\alpha}$, $\sigma_{s_i,r_j}^2=\eta D_{s_i,r_j}^{-\alpha}$ and $\sigma_{r_j,d}^2=\eta D_{r_j,d}^{-\alpha}$ respectively where $D_{s_i,d}$, $D_{s_i,r_j}$ and $D_{r_j,d}$ are the respective link distances, $\alpha$ is path loss exponent, $\eta$ is propagation environment dependent constant. The additive white Gaussian noise (AWGN) power at the relay nodes and BS are $N_0$. In cooperative mode, the source node $s_i$ broadcasts messages with transmit power $P_{s_i}$ in the first phase of cooperation. During the second phase, a selected relay node $r_j$, transmits messages with transmit power $P_{r_j}$ in AF mode \cite{coop_protocols_laneman}. Upon receiving two copies of the message, the BS uses maximal ratio combining (MRC) to detect the transmitted symbols. Thus SER at BS at moderate to high average link signal-to-noise ratio (SNR) is  \cite{e_to_e_ber_AF}
\begin{equation}
\label{eq_2_1}
p_{e} = \frac{3}{4K^2 \bar{\gamma}_{s_i,d}} \left( \frac{1}{\bar{\gamma}_{s_i,r_j}}+\frac{1}{\bar{\gamma}_{r_j,d}}\right)
\end{equation}
where $K$ is a modulation index dependent parameter, $\bar{\gamma}_{s_i,d}$, $\bar{\gamma}_{s_i,r_j}$ and $\bar{\gamma}_{r_j,d}$ are the average SNR of the source node $s_i$-to-BS, source node $s_i$-to-relay node $r_j$ and relay node $r_j$-to-BS links, respectively. The average SNR terms can be expressed as $\bar{\gamma}_{s_i,d}=G_0 P_{s_i}\sigma_{s_i,d}^2/N_0$, $\bar{\gamma}_{s_i,r_j}=G_0 P_{s_i}\sigma_{s_i,r_j}^2/N_0$ and $\bar{\gamma}_{r_j,d}=G_0 P_{r_j}\sigma_{r_j,d}^2/N_0$ where $G_0$ as the power gain factor at reference distance $1$ m. Substituting the values of average SNR terms in (\ref{eq_2_1}), end-to-end SER can be expressed as
\begin{equation}
\label{eq_2_2}
p_{e}=\frac{A_{s_i,r_j}}{P_{s_i}^2 }+\frac{B_{s_i,r_j}}{P_{s_i}P_{r_j}}
\end{equation}
where $A_{s_i,r_j}=3N_0^2/4K^2G_0^2\sigma_{s_i,d}^2\sigma_{s_i,r_j}^2$ and $B_{s_i,r_j}=3N_0^2/4K^2G_0^2\sigma_{s_i,d}^2\sigma_{r_j,d}^2$.

\subsection{Problem Formulation}
For cooperative network, the lifetime definition considered in literature are: ‘end-to-end QoS’ lifetime \cite{coop_pa_relay_select_outage_constrained2, prediction_lifetime_coop_rate, lifetime_cooperative_rate_constrained, lifetime_coop_multiuser_rly_select}, ‘first node death’ lifetime \cite{coop_pa_1_rs_rate_constr, coop_pa_1_rs_rate_constrained2}. The ‘end-to-end QoS’ lifetime for multiple-user network is defined as the time interval within which end-to-end QoS of all the source messages is maintained through their respective relay nodes \cite{lifetime_coop_multiuser_rly_select}. Since we consider both source and relay nodes as energy limited, according to this definition the network expires if any of the source or relay nodes fails to transmit data with available residual energy. Therefore when all nodes are energy limited, ‘end-to-end QoS’ lifetime is same as ‘first node death’ lifetime which is defined as the time interval until first node in the network is depleted of energy \cite{coop_pa_1_rs_rate_constr, coop_pa_1_rs_rate_constrained2}. 
Throughout the paper we use ‘first node death’ lifetime as network lifetime definition. This definition is useful in a network where all the nodes are equally important, e.g. surveillance applications. The GLM policy aims to maximize time interval of first node death occur in the network according to the current network state, i.e., residual energy level, location of the nodes. Let $T_{s_i}=E_{s_i}/P_{s_i}$, $T_{r_j}=E_{r_j}/P_{r_j}$ are the lifetime of the source node $s_i$ and  relay node $r_j$ with residual energies $E_{s_i}$ and $E_{r_j}$ respectively. Therefore lifetime of the source relay pair $(s_i,r_j)$ can be expressed as 
\begin{align}
\label{eq_2_3}
T_{g,(s_i,r_j)}= \min \left( T_{s_i}, T_{r_j} \right)
\end{align} 
Let $\Pi$ denotes the set of all possible source relay assignments of the source set $S$ and relay set $R$ such that every set $\pi \in \Pi$ is the  source relay assignments that contains $M$ disjoint source relay pair. For example, with $S=\{s_1, s_2\}$ and $R=\{r_1, r_2\}$ we have two different source relay assignments partition $\{(s_1,r_1),(s_2,r_2)\}$, $\{(s_1,r_2),(s_2,r_1)\}$ and
$\Pi=\{\{(s_1,r_1),(s_2,r_2)\}, \{(s_1,r_2),(s_2,r_1)\}\}$. Then the relay assignment and power allocation problem for GLM strategy can be formulated as
\begin{align}
\label{eq_2_5}
\nonumber \max_{\pi \in \Pi, \matr{P}_{(s_i,r_j)}}& \min_{(s_i,r_j)\in \pi}  \;\;\;  T_{g,(s_i,r_j)} \\  \nonumber 
s.t.&  \quad \frac{A_{s_i,r_j}}{P_{s_i}^2 }+\frac{B_{s_i,r_j}}{P_{s_i}P_{r_j}} \leq p_{th}^{s_i} ,  \;\; \forall (s_i,r_j)\in \pi
\\ &  \quad \matr{P}_{(s_i,r_j)}> \matr{0},  \;\; \forall (s_i,r_j)\in \pi
\end{align}
where $\matr{P}_{(s_i,r_j)}=[P_{s_i}, P_{r_j}]$ is the transmit power vector of source relay pair $(s_i,r_j)$, $p_{th}^{s_i}$ is the end-to-end SER constraint for source node $s_i$. Note that relay assignment and power allocation according to GLM policy is optimal for network lifetime maximization if the allocated transmit power and relay selection remains fixed and not updated till the network expires. To compare with network lifetime performance of GLM policy, we propose another residual energy aware relay selection and power allocation policy referred as MWTP policy. In MWTP policy, we aim to minimize residual energy weighted total power, i.e.,
\begin{align}
\label{eq_2_4}
\nonumber \min_{\pi \in \Pi, \matr{P}_{(s_i,r_j)}} & \quad \sum_{(s_i,r_j)\in \pi} \frac{P_{s_i}}{E_{s_i}}+\frac{P_{r_j}}{E_{r_j}} \\  \nonumber 
s.t.&  \quad \frac{A_{s_i,r_j}}{P_{s_i}^2 }+\frac{B_{s_i,r_j}}{P_{s_i}P_{r_j}} \leq p_{th}^{s_i}, \;\; \forall (s_i,r_j)\in \pi
\\ &  \quad \matr{P}_{(s_i,r_j)}> \matr{0},  \;\; \forall (s_i,r_j)\in \pi
\end{align}
 
\section{Power Allocation for a Fixed Pair}
\label{sec_power_allocation}
 \subsection{GLM policy}
 \label{sec_pa_glm}
The power allocation problem for GLM policy is gven by
\begin{align}
\label{eq_3b_1}
\nonumber \max & \;\;\;  T_{g,(s_i,r_j)}  \\  
s.t.&  \quad \frac{A_{s_i,r_j}}{P_{s_i}^2 }+\frac{B_{s_i,r_j}}{P_{s_i}P_{r_j}}\leq p_{th}^{s_i},
\;\; P_{s_i}> 0, P_{r_j}> 0.
\end{align}
By replacing $T_{g,(s_i,r_j)}=1/V$ and using (\ref{eq_2_3}), the power allocation problem becomes
\begin{align}
\label{eq_3b_2}
\nonumber \min& \;\;\;  V
\\ \nonumber s.t.& \; \;\; P_{s_i} \leq E_{s_i} V, \; P_{r_j} \leq E_{r_j} V, \\  &\; \;\;  \frac{A_{s_i,r_j}}{P_{s_i}^2 }+\frac{B_{s_i,r_j}}{P_{s_i}P_{r_j}} \leq p_{th}^{s_i}, \;\; P_{s_i}> 0, P_{r_j}> 0.
\end{align}
The optimization problem is convex and has a
unique optimal solution \cite{boyd}. From the Karush-Kuhn-Tucker (KKT) conditions we have 
\begin{equation}
\label{eq_3b_3}
P_{s_i} =E_{s_i} V, \; P_{r_j} =E_{r_j} V,
\end{equation}
\begin{equation}
\label{eq_3b_5}
\quad \frac{A_{s_i,r_j}}{P_{s_i}^2 }+\frac{B_{s_i,r_j}}{P_{s_i}P_{r_j}}= p_{th}^{s_i}.
\end{equation}
Hence, using (\ref{eq_3b_3}), (\ref{eq_3b_5}), the optimal transmit power of $s_i$ and $r_j$ becomes
\begin{equation}
\label{eq_3b_5a}
P_{s_i}^{glm}=\sqrt{\frac{A_{s_i,r_j}E_{r_j}+B_{s_i,r_j}E_{s_i} }{ E_{r_j} p_{th}^{s_i}}}, \; P_{r_j}^{glm}=P_{s_i}^{glm} \frac{E_{r_j}}{E_{s_i}}
\end{equation}
and optimal lifetime of the source relay pair $(s_i,r_j)$ is
\begin{align}
T_{g,(s_i,r_j)}^{glm}=\frac{E_{s_i}}{P_{s_i}^{glm}}=\frac{E_{r_j}}{P_{r_j}^{glm}}
\end{align}
Hence, lifetime of the pair is maximized when both source and relay nodes die at same time. Let weight of source relay pair $(s_i,r_j)$ for GLM policy is
\begin{equation}
\label{eq_3b_6}
\omega_{(s_i,r_j)}=\frac{1}{T_{g,(s_i,r_j)}^{glm}}=\sqrt{\frac{A_{s_i,r_j}E_{r_j}+B_{s_i,r_j}E_{s_i} }{ E_{s_i}^2 E_{r_j} p_{th}^{s_i}}}.
\end{equation}

 \subsection{MWTP policy} 
  \label{sec_pa_mwtp}
The power allocation problem for MWTP policy is
\begin{align}
\label{eq_3c_1}
\nonumber \min& \quad \frac{P_{s_i}}{E_{s_i}}+\frac{P_{r_j}}{E_{r_j}} \\  
s.t.&  \quad \frac{A_{s_i,r_j}}{P_{s_i}^2 }+\frac{B_{s_i,r_j}}{P_{s_i}P_{r_j}} \leq p_{th}^{s_i}, \;\; P_{s_i}> 0, P_{r_j}> 0.
\end{align}
Since the objective function is linear and the constraint is convex function, the optimization problem is convex problem and has a
unique optimal solution \cite{boyd}. Since end-to-end SER decreases with increase of $P_{s_i}$, $P_{r_j}$, to minimize the objective function,  $P_{s_i}$, $P_{r_j}$ must satisfy 
\begin{align}
\frac{A_{s_i,r_j}}{P_{s_i}^2 }+\frac{B_{s_i,r_j}}{P_{s_i}P_{r_j}} = p_{th}^{s_i}.
\end{align}
Therefore $P_{r_j}$ can be expressed in terms of $P_{s_i}$ as
\begin{align}
 P_{r_j}=\frac{B_{s_i,r_j} P_{s_i}}{p_{th}^{s_i} P_{s_i}^2-A_{s_i,r_j}}\triangleq f\left( P_{s_i} \right)
\end{align}
The optimal transmit power can be obtained by setting
the derivative on $P_{s_i}/E_{s_i}+f( P_{s_i} )/E_{r_j}$ to be zero, as
\begin{align}
\label{eq_3c_7}
\nonumber P_{s_i}^{mwtp}&=\sqrt{\frac{A_{s_i,r_j} \left( 2+C_{s_i,r_j}+\sqrt{C_{s_i,r_j}^2+8C_{s_i,r_j} } \right)}{2p_{th}^{s_i}}}, \\ P_{r_j}^{mwtp}&=f\left( P_{s_i}^{mwtp} \right). 
\end{align}
where $C_{s_i,r_j}=B_{s_i,r_j}E_{s_i}/A_{s_i,r_j}E_{r_j}$. Then weight of source relay pair $(s_i, r_j)$ for MWTP policy can be expressed as
\begin{align}
\label{eq_3c_8}
\omega_{(s_i,r_j)}=\frac{P_{s_i}^{mwtp}}{{E_{s_i}}}+\frac{P_{r_j}^{mwtp}}{E_{r_j}}.
\end{align}

 \section{Relay Selection Schemes}
 \label{sec_relay_selection}
Consider first the relay assignment and power allocation problem for MWTP ploicy as given in (\ref{eq_2_4}). With use of weight $\omega_{(s_i,r_j)}$ assigned to each source relay pair $(s_i,r_j)$ according to the optimal power allocation solution, the optimization problem of MWTP policy as given in (\ref{eq_2_4}) reduces to
\begin{align}
\label{eq_4a_3}
\min_{\pi \in \Pi} & \; \sum_{(s_i,r_j)\in \pi}    \omega_{(s_i,r_j)}  
\end{align}
where $\omega_{(s_i,r_j)}$ is defined in  (\ref{eq_3c_8}) for  MWTP policy.

With weight assigned to each source relay pair $(s_i,r_j)$ in (\ref{eq_3b_6}) according to the optimal power allocation solution for GLM policy, the optimization problem for GLM policy given in (\ref{eq_2_5}), becomes
 \begin{align}
\label{eq_4a_8}
 \min_{\pi \in \Pi} & \; \max_{(s_i,r_j)\in \pi}   \omega_{(s_i,r_j)}  
\end{align} 
Both the relay assignment problems given in (\ref{eq_4a_3}) and (\ref{eq_4a_8}), can be solved with the exhaustive search over all possible relay assignments with complexity $\mathcal{O}(MN^{M}) \sim \mathcal{O}(M^2N^{M})$ or exhaustive search over all possible source priority vector with complexity of $\mathcal{O}(NM^{M+1})$ as proposed in \cite{lifetime_coop_multiuser_rly_select}. In this section we propose graph theoretic approach to solve optimal relay assignment problem with lower complexity compared to these strategies. We further propose suboptimal relay selection method with lower complexity.

 \subsection{Optimal Relay Selection with Bipartite Matching Approach}
 \label{sec_bipartite_matching}
Before we proceed, we review some preliminary concepts of bipartite graph theory matching \cite{combinatorial_optimization,hungarian, bottleneck_matching}. A graph $G$ comprising of vertex set $\mathcal{V}_G$ and edge set $\mathcal{E}_G$ is bipartite if $\mathcal{V}_G$ can be partitioned into two sets $\mathcal{V}_G^1$ and $\mathcal{V}_G^2$ (the bipartition) such that every edge in $\mathcal{E}_G$ connects a vertex in $\mathcal{V}_G^1$ to one in $\mathcal{V}_G^2$. A complete bipartite graph is a bipartite graph where every vertex of $\mathcal{V}_G^1$ is connected by an edge to every vertex of $\mathcal{V}_G^2$, i.e., $|\mathcal{E}_G|=|\mathcal{V}_G^1||\mathcal{V}_G^2|$ where $|\mathcal{E}_G|$ is the number of edges and $|\mathcal{V}_G^1|$, $|\mathcal{V}_G^2|$ are the number of vertices in $\mathcal{V}_G^1$, $\mathcal{V}_G^2$ of the graph, respectively. If the two sets of vertices have the same cardinality, i.e., $|\mathcal{V}_G^1|=|\mathcal{V}_G^2|=|\mathcal{V}_G|/2$,  then the bipartite graph is a balanced bipartite graph. A matching in a graph $G$ is a subset of $\mathcal{E}_G$ such that every vertex $v \in \mathcal{V}_G$ is incident to at most one edge of the matching. Maximum matching $M^{\ast}$ in $G$ is a matching that contains the largest possible number of edges. The maximum matching $M^{\ast}$ is perfect matching if each vertex $v \in \mathcal{V}_G$ belongs to an edge in $M^{\ast}$. Clearly, for a balanced bipartite graph, a maximum matching $M^{\ast}$ is perfect matching if $|M^{\ast}|=|\mathcal{V}_G|/2$. Also note that, a balanced, complete bipartite graph always has perfect matching.


Now, if the network is represented as a complete bipartite  graph $G$ such that $\mathcal{V}_G^1=\{s_1, ..,s_M\}$ and $\mathcal{V}_G^2=\{r_1, ..,r_N\}$ are the vertex sets of $G$ and weight of each edge $(s_i,r_j)$ between vertices $s_i$ and $r_j$, $i \in 1,.., M$, $j \in 1,.., N$ is $\omega_{(s_i,r_j)}$ as given in (\ref{eq_3c_8}), the relay assignment problem in (\ref{eq_4a_3}) can be described as finding maximum matching in the graph $G$ such that sum of edge weights in the matching has minimum value. This problem is known as minimum weighted matching (MWM) problem in a bipartite graph.  
The MWM algorithm in \cite{hungarian} (also known as Hungarian algorithm) can solve the MWM problem optimally for a balanced bipartite graph. Therefore if $M<N$, $N-M$ dummy vertices that have zero edge weights with each vertex in $\mathcal{V}_G^2$, require to be added in $\mathcal{V}_G^1$ of the graph $G$ \cite{combinatorial_optimization}. Then, relay assignment problem in (\ref{eq_4a_3}) can be  solved in time $\mathcal{O}(N^{3} )$ using MWM algorithm on the graph \cite{hungarian}.

The bottleneck matching (BM) problem in a bipartite graph is defined as finding maximum matching in the graph such that the largest edge weight of the matching is as small as possible. If the network is represented as a complete bipartite graph $G$ such that $\mathcal{V}_G^1=\{s_1, ..,s_M\}$ and $\mathcal{V}_G^2=\{r_1, ..,r_N\}$ are the vertex sets of $G$ and  weight of each edge $(s_i,r_j)$ is $\omega_{(s_i,r_j)}$ as given in (\ref{eq_3b_6}), the relay assignment problem in (\ref{eq_4a_8}) can be described as BM problem of the graph $G$. The BM problem for balanced bipartite graph has been optimally solved in \cite{bottleneck_matching}. Therefore if $M<N$, $N-M$ dummy vertices that have zero edge weights with each vertex in $\mathcal{V}_G^2$, are added in $\mathcal{V}_G^1$ of $G$ \cite{combinatorial_optimization}. Then, the relay assignment problem in (\ref{eq_4a_8}) can be solved in time $\mathcal{O}(N^{2.5} )$ by applying BM algorithm on the graph \cite{bottleneck_matching}.

\begin{figure}
\centering
\includegraphics[width=0.35\textwidth]{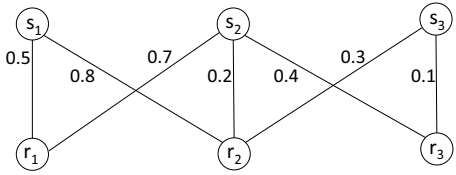}
\vspace*{-0.4cm}
\caption{An example of weighted bipartite graph with unique bottleneck edge.}
\label{fig0_1}
\vspace{-0.3cm}
\end{figure}

The bottleneck matching in a bipartite graph is not necessarily unique. For example, consider the bipartite graph with edge weights as shown in the  Fig.\ref{fig0_1}. The graph has two different possible bottleneck matching $\mathcal{B}_1=\{(s_1,r_1), (s_2,r_2), (s_3,r_3)\}$ and $\mathcal{B}_2=\{(s_1,r_1), (s_2,r_3), (s_3,r_2)\}$. Note that, both the bottleneck matching $\mathcal{B}_1$ and $\mathcal{B}_2$, has same bottleneck edge $(s_1,r_1)$, with weight $\omega_{(s_1,r_1)}=0.5$. 
\begin{definition}
An edge $e$ of a bipartite graph $G$ is unique  bottleneck edge if the edge $e$ corresponds to the maximum weight in every bottleneck matching of the graph $G$.
\end{definition} 
Therefore $(s_1,r_1)$ is the unique bottleneck edge of the graph. Now the question is, if relay selection according to one bottleneck matching will have same network lifetime performance always compared to relay selection according to another bottleneck matching in the network. To explain this, consider the example of the network corresponding to the graph shown in the  Fig.\ref{fig0_1}. The network has three source nodes and three relay nodes. The weight $\omega_{(s_i,r_j)}$ of each edge $(s_i,r_j)$, $i,j \in \{1,2,3\}$ shown in Fig.\ref{fig0_1} is actually weight of source relay pair $(s_i,r_j)$, calculated using (\ref{eq_3b_6}) at time $t=0$. Let the initial energies of  each source node and relay node is $E_{s_i}=E_{r_j}=1 J$. Therefore, average transmit power of nodes $s_i$, $r_j$ is $\omega_{(s_i,r_j)}$ and lifetime of nodes $s_i$, $r_j$ is $1/\omega_{(s_i,r_j)}$ when $s_i$ is paired with $r_j$. Therefore  lifetime of the network is same as lifetime of source relay pair $(s_1,r_1)$ which is $1/\omega_{(s_1,r_1)}=2 $ seconds when relay selection is according to the bottleneck matching $\mathcal{B}_1$ or $\mathcal{B}_2$. However the power consumption of source nodes $s_2, s_3$ relay nodes $r_2,r_3$ is lower if the relay selection at $t=0$ is according to  $\mathcal{B}_1$ rather than $\mathcal{B}_2$. 
Therefore at $t=t'$ where $0<t'<2$, the residual energy of source nodes $s_2,s_3$ relay nodes $r_2,r_3$ is higher if the initial relay selection at $t=0$ was according to $\mathcal{B}_1$ instead of $\mathcal{B}_2$. Now if the edge weights are updated again at time $t=t'$, (to update the relay selection according to the residual energy state in the network at $t=t'$), the weight of all edges except $(s_1,r_1)$ is higher if the relay selection at $t=0$, was according to $\mathcal{B}_1$ rather than $\mathcal{B}_2$. Thus, if the initial relay selection at $t=0$ was according to $\mathcal{B}_1$ instead of $\mathcal{B}_2$, there is a possibility of obtaining a bottleneck matching with lower bottleneck weight (i.e., higher network lifetime) when relay selection is updated at time $t=t'$. 
 \begin{definition}
 The minimum bottleneck matching of a bipartite graph $G$ is defined as  maximum matching in $G$ such that largest edge weight, second largest edge weight, .., $\mathcal{M}$ th largest edge weight of the matching is as small as possible where $\mathcal{M}$ is number of edges in the  maximum matching. 
\end{definition} 
Note that choice of the edge corresponding to $i$th largest edge weight of the matching with minimum possible value where $i \in 2,.., \mathcal{M}$, depends upon previously selected edges
corresponding to the minimum possible largest edge weight, second largest edge weight, ..., $(i-1)$th largest edge weight of the matching. For example, in Fig.\ref{fig0_1}, among all possible maximum matching in the graph, $(s_1,r_1)$ is the edge corresponding to minimum possible largest edge weight, i.e., the bottleneck edge. Then, among the maximum matching in which $(s_1,r_1)$ is the edge with largest edge weight (i.e., $\mathcal{B}_1$ and $\mathcal{B}_2$), second largest edge weight in $\mathcal{B}_1$ is smaller than of $\mathcal{B}_2$. Therefore, $\mathcal{B}_1$ is the minimum bottleneck matching of the graph. As discussed above, the relay selection with minimum bottleneck matching may have better network lifetime performance compared to other bottleneck matching when the relay selection and power allocation are updated periodically. However, the traditional bottleneck matching algorithm \cite{bottleneck_matching} do not ensure to find minimum bottleneck matching but finds any of the bottleneck matching for the graph. Inspired by all the above observations, we propose a novel   algorithm for relay selection as shown in Algorithm \ref{algo_mbm}, which is referred to as the minimum bottleneck matching (MBM) algorithm. The algorithm runs for maximum $M$ iterations for the graph $G$. At each iteration, a bottleneck matching with corresponding bottleneck edge is obtained. Then, if the bottleneck edge is unique bottleneck edge, the bottleneck edge is stored and the graph is updated by removing vertices corresponding to the bottleneck edge. If the graph do not have unique bottleneck edge during any of the iterations, the algorithm terminates with output bottleneck matching $\mathcal{B}_{o}$ as bottleneck matching obtained in the first iteration and acknowledgement metric $I=0$, which states that the output bottleneck matching $\mathcal{B}_{o}$ is not necessarily a minimum bottleneck matching. This is because, if the obtained bottleneck edge at any of the iteration is not unique bottleneck edge, it can not be decided which of the bottleneck edges is an edge of minimum bottleneck   matching and finding minimum bottleneck matching is difficult in such case. If the graph has unique bottleneck edge at each iteration, after $M$ successful iterations, the algorithm  returns the minimum bottleneck matching which is a set of all unique bottleneck edges obtained in each iterations, and the acknowledgement metric $I=1$ stating that the output bottleneck $\mathcal{B}_{o}$ matching is the minimum bottleneck matching. To test if a bottleneck edge corresponding to a bottleneck matching is unique bottleneck edge in the graph, the following steps are used in the algorithm. Firstly, remove the bottleneck edge and the edges with weight greater than bottleneck weight from the graph and check if the graph has a perfect matching. If the perfect matching exists, then the bottleneck edge is not an unique bottleneck edge, otherwise it is unique bottleneck edge.  This is because, the bipartite graph has perfect matching since it is balanced, complete and the maximum weight in every perfect matching of the graph is greater than or equal to the bottleneck weight. Now since the bottleneck edge and the edges with weight greater than bottleneck weight are removed, the perfect matching exist only if maximum weight of the perfect matching is equal to bottleneck weight. Therefore the perfect matching is actually a bottleneck matching of the original graph with different bottleneck edge and the removed bottleneck edge is not unique bottleneck edge. Note that, in a graph corresponding to a random network, weight of two or more different edges  to be exactly same is rare as we have tested  in the simulations. In case of a graph with non-identical weight edges, minimum bottleneck matching is always available and MBM algorithm can successfully find it  since the bottleneck edge obtained for the graph at each iteration is unique bottleneck edge.  The MBM algorithm at each iteration run the bottleneck matching algorithm in time $\mathcal{O}(N^{2.5} )$ \cite{bottleneck_matching} and maximum matching algorithm in time $\mathcal{O}(N^{w} )$ where $w<2.38$ \cite{max_matching_general_graph}. Therefore, time complexity of the MBM algorithm is $\mathcal{O}(MN^{2.5} )$.

\begin{algorithm}
\begin{algorithmic}[1]
\State Construct graph $G$ with vertex sets $\mathcal{V}_G^1=\{s_1, ..,s_N\}$, $\mathcal{V}_G^2=\{r_1, ..,r_N\}$. Weight of each  edge between vertices $s_i$, $r_j$, $i \in 1,.., M$, $j \in 1,.., N$, $\omega_{(s_i,r_j)}$ is according to (\ref{eq_3b_6}) and weight of  each edge between vertices $s_l$, $l \in M+1,.., N$ (i.e., dummy vertex) and $r_j$, $\omega_{(s_l,r_j)}=0$. 
\State Initialize acknowledgement metric $I=1$, $H=G$, a bottleneck matching of $G$ is $\mathcal{B}_G$ using \cite{bottleneck_matching} and $\mathcal{B}_o=\emptyset$.
\For {$i=1:M$}
\State Find bottleneck matching $\mathcal{B}_H$ of the graph $H$ and corresponding bottleneck edge $b_H$ using \cite{bottleneck_matching}.
\State $H'=H-\{b_H,\mathcal{E}'\}$ where $\mathcal{E}'=\{e \in \mathcal{E}_H|\omega_{e}>\omega_{b_H}\}$ is set of edges with weight greater than  bottleneck weight.
\State Find maximum matching $M^{\ast}$ of the graph $H'$
\If {$|M^{\ast}|= |\mathcal{V}_H|/2$} 
\State $\mathcal{B}_{o}=\mathcal{B}_G$
\State $I=0$
\State Break
\EndIf
\State $\mathcal{B}_{o}=\mathcal{B}_{o} \cup b_H$
\State $H=H-v_{b_H}$ where $v_{b_H}$ is vertex set of the edge $b_H$. 
\EndFor
\end{algorithmic}
\caption{Minimum bottleneck matching algorithm}
\label{algo_mbm}
\end{algorithm}


 \begin{figure}
\centering
\includegraphics[width=0.48\textwidth]{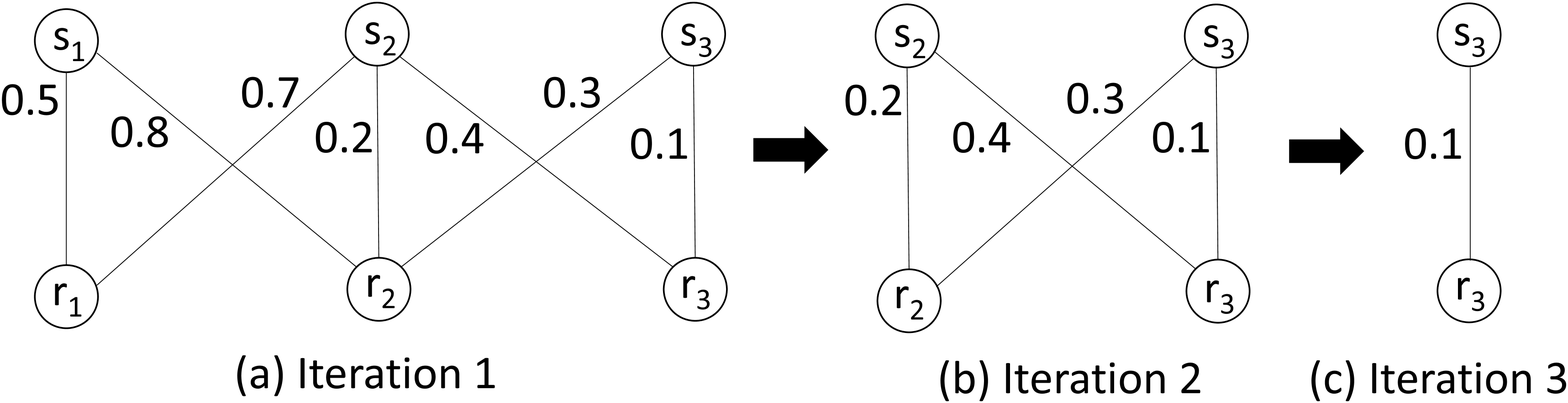}
\vspace*{-0.4cm}
\caption{An example of graph updation with iterations of MBM algorithm.}
\label{fig0_2}
\vspace{-0.3cm}
\end{figure}

The algorithm can be further explained with the example of how it works for the graph given in Fig.\ref{fig0_1}.  The graph at each iteration of the algorithm has been shown in Fig.\ref{fig0_2}. At the first iteration, the bottleneck matching is either $\mathcal{B}_1$ or $\mathcal{B}_2$. Therefore in either case bottleneck edge is $(s_1,r_1)$ is to be tested if it is unique bottleneck edge. For this purpose, the edges $(s_1,r_1)$, $(s_1,r_2)$ and $(s_2,r_1)$ are removed from the graph and maximum matching is obtained on the resultant graph. Now, the vertices $s_1$, $r_1$ of the resultant graph are not connected to any edge. Therefore, the maximum matching on the resultant graph is not a perfect matching and edge $(s_1,r_1)$ is unique bottleneck edge. The edge $(s_1,r_1)$ is stored and the graph is updated by removing the vertices $s_1$, $r_1$ (and corresponding edges). Using similar steps at second and third iteration, the unique bottleneck edges for the graph for the corresponding iteration are obtained as $(s_2,r_2)$ and $(s_3,r_3)$, respectively. Therefore the minimum bottleneck  matching $\mathcal{B}_1$ is obtained finally after three successful iterations.

\subsection{Suboptimal Relay Selection}
\label{sec_suboptimal_algo}
 Now we propose suboptimal relay selection (SRS) algorithm for the multiple-user network. The low complexity relay selection algorithm in multiple-user network can be solved in two steps: firstly to find a low complexity suboptimal source priority metric, then selecting best relay nodes sequentially according to the order of source priority metric \cite{lifetime_coop_multiuser_rly_select, coop_multiuser_performance_min_snr_rly_select}. Source priority can be found based on weights of source relay pairs. For example, higher priority is given to the source node that has larger weight with its best relay  \cite{coop_multiuser_performance_min_snr_rly_select} or worst relay \cite{lifetime_coop_multiuser_rly_select}. However in these cases, complexity of relay selection are still high. In the proposed SRS algorithm, source priority is decided based on source node average channel condition (distance from the BS) and its residual energy level. To decide the source priority, we use a priority metric $\wp_{s_i}$. The source node $s_i$ with higher value of $\wp_{s_i}$ has higher source priority. For GLM and MWTP policies we propose the source priority metric as
  \begin{align}
   \label{eq_4b_2}
 \wp_{s_i}=\frac{1}{E_{s_i}\sigma_{s_i,d}^2}, \;\; s_i \in S
 \end{align}
i.e., a source node with lower residual energy or located far from BS (i.e., higher power consumption) is given higher priority to select its best relay since the node is more likely to die out early.

\begin{algorithm}
\begin{algorithmic}[1]
\State Let $S=\{s_1,s_2,.., s_M\}$ and $R=\{r_1,r_2,.., r_N\}$. 
\State Compute the priority metric $\wp_{s_i}$ for   source node $s_i \in S$.
\State Initialize $m_{s_i}=-1$, for $s_i \in S$.
\While {$|S| > 0$}
\State $s_{i}^{\ast}=\argmin_{s_{i} \in S} \wp_{s_i}$
\State $S=S \setminus s_{i}^{\ast}$
\ForAll {$r_j \in R$}
\State Calculate $\omega_{s_{i}^{\ast},r_j}$ according to the policy.
\EndFor
\State $r_{j}^{\ast}=\argmin_{r_j \in R} \omega_{s_{i}^{\ast},r_j}$
\State $m_{s_{i}^{\ast}}=r_{j}^{\ast}$
\State $R=R \setminus r_{j}^{\ast}$
\EndWhile
\end{algorithmic}
\caption{Suboptimal relay selection (SRS) algorithm}
\label{algo1}
\end{algorithm}

The proposed SRS algorithm is described in Algorithm \ref{algo1}. According to this algorithm, a source node is chosen from the unpaired source set according to source priority metric and then the relay with minimum source relay weight from the unpaired relay set is found and paired with the source node. This process continues repeatedly until each  source node is allocated a relay node. For each source node $s_i$, $m_{s_i}$ denotes its assigned relay. Compared to the suboptimal algorithm proposed in \cite{lifetime_coop_multiuser_rly_select} with complexity  $\mathcal{O}(NM^2)$ which is quadratic in number of source node, our proposed algorithm complexity behaves as $\mathcal{O}(MN)$, linear in the number of source node and relay nodes. This is because of the fact that in SRS algorithm, source priority is found with lower complexity.

Therefore, for GLM policy, we have three relay selection and power allocation strategies: GLM BM, GLM MBM and GLM SRS, with time complexity $\mathcal{O}(N^{2.5} )$, $\mathcal{O}(MN^{2.5})$ and $\mathcal{O}(MN )$, respectively. The GLM BM, GLM MBM and GLM SRS strategies use bottleneck matching algorithm  \cite{bottleneck_matching},  Algorithm \ref{algo_mbm} and Algorithm \ref{algo1}, respectively with weight of each source relay pair according to (\ref{eq_3b_6}) to obtain the relay assignment in the network and assign the transmit power to each  selected source relay pair according to the power allocation solution in (\ref{eq_3b_5a}). Similarly we have MWTP MWM and MWTP SRS strategies with time complexity $\mathcal{O}(N^{3} )$ and $\mathcal{O}(MN )$,  respectively, for relay selection and power allocation for MWTP policy. Note that, GLM BM and GLM MBM strategies optimally solve (\ref{eq_2_5}). Hence these strategies are optimal for network lifetime maximization if relay selection and power allocation is not updated before the network expires.

\section{PERFORMANCE EVALUATION}
\label{sec_results}
In this section, we present the simulation results to evaluate the network lifetime performance of the proposed strategies. Consider a network with six source nodes. Let the source nodes and relay nodes are distributed randomly in a square area of size 100 m$\times$100 m and BS is located at ($50$, $50$). Each source node and relay  node has initial energy $10$ J. The system parameters for all simulations are data rate $10$ Kbps, $G_0$=-70 dB, $N_0$=-134 dBm, $\alpha$=3.5, $\eta=1$, $K$=2 i.e., BPSK modulation. SER constraint for each source node is $10^{-4}$. The network lifetime is measured in terms of total number of data packet received at the BS since this is equivalent to the time the network is operating. The results are averaged over $300$ randomly generated network topologies.

\begin{figure}
\centering
\includegraphics[width=0.48\textwidth]{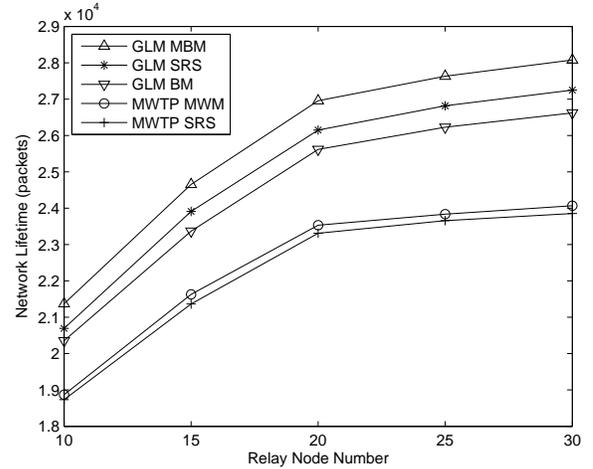}
\vspace*{-0.4cm}
\caption{Network lifetime versus relay node number in $100m\times100m$ network with six source nodes.}
\label{fig1}
\vspace{-0.3cm}
\end{figure}

Fig.\ref{fig1} depicts the lifetime performance of the proposed strategies when number of relay nodes in the network varies from $10$ to $30$ and SER constraint is $10^{-4}$. The relay selection and power allocation for each strategy is periodically updated after all six source nodes transmit $10$ packets. Therefore the update interval is $T_{u}=60$ packets. As relay number increases, the network lifetime increases for all the strategies. This is because adding more relay nodes in the network increases total energy of the network along with the increase of availability of more efficient relay nodes. The GLM policy based strategies perform better than MWTP policy based strategies and GLM MBM achieves maximum network lifetime. The GLM SRS and MWTP SRS perform close to GLM MBM and MWTP MWM strategies, respectively. The network lifetime of  GLM BM is lower compared to the GLM MBM and GLM SRS since GLM BM is not power efficient when $T_{u}$ is low. With $20$ relay nodes, GLM MBM has $1.03$ and $1.05$ times more network lifetime than GLM SRS and GLM BM, respectively while GLM SRS achieves $1.11$, $1.12$ times more network lifetime than MWTP MWM, MWTP SRS, respectively.

\begin{figure}
\centering
\includegraphics[width=0.48\textwidth]{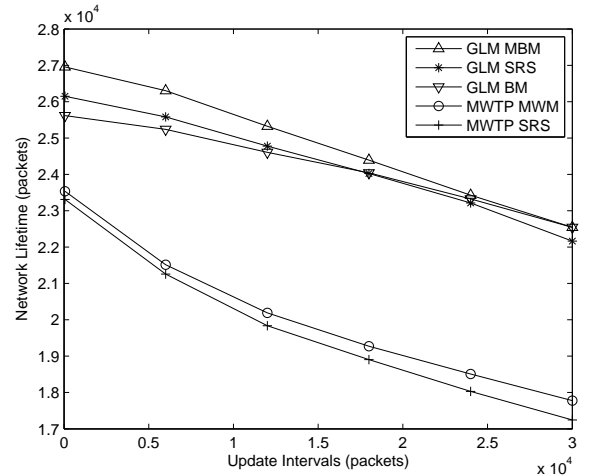}
\vspace*{-0.4cm}
\caption{Network lifetime versus update interval in $100m\times100m$ network with six source nodes and twenty relay nodes.}
\label{fig2}
\vspace{-0.3cm}
\end{figure}

In Fig.\ref{fig2}, we compare the lifetime performance of these strategies for different update interval $T_{u}$ with six source nodes and twenty relay nodes. As $T_{u}$ increases, the network lifetime decreases for all strategies. The reason is that with the increase of $T_{u}$, update of the relay nodes become more infrequent and few relay nodes carry heavier traffic compared to other which results in early die out of these relay nodes. It can be observed that for low $T_{u}$, the GLM BM strategy has lower network lifetime performance compared to GLM MBM as well as GLM SRS. A cross over occurs at $T_{u}=1.7 \times 10^4 $ packets, beyond which the GLM BM strategy has better network lifetime performance compared to the GLM SRS strategy. As discussed in Section \ref{sec_relay_selection}, if the the network expires before an updation of relay selection and power allocation, GLM BM and GLM MBM are optimal for lifetime maximization and these strategies have same network lifetime performance. This situation is shown in Fig. \ref{fig2} for $T_{u}=3 \times10^4$ packets which is higher than network lifetime for all the strategies.

\begin{figure}
\centering
\includegraphics[width=0.48\textwidth]{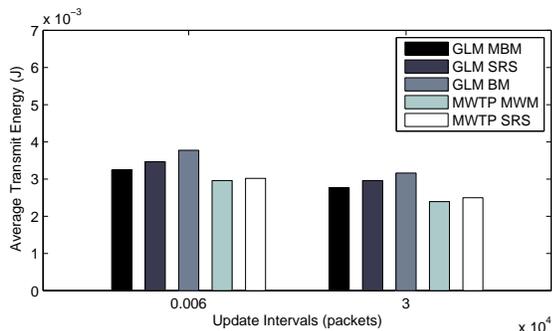}
\vspace*{-0.4cm}
\caption{Average transmit energy per packet for different update interval in $100m\times100m$ network with six source nodes and twenty relay nodes.}
\label{fig3a}
\vspace{-0.3cm}
\end{figure}

\begin{figure}
\centering
\includegraphics[width=0.48\textwidth]{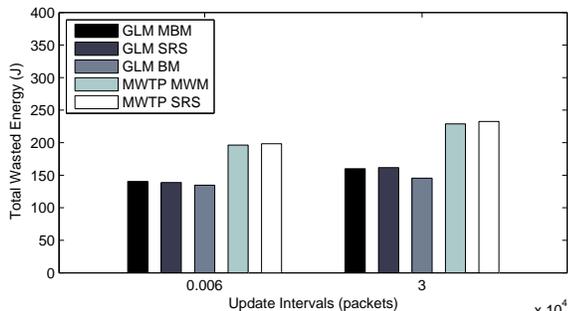}
\vspace*{-0.4cm}
\caption{Total wasted energy for different update interval in $100m\times100m$ network with six source nodes and twenty relay nodes.}
\label{fig3b}
\vspace{-0.3cm}
\end{figure}

To further understand the network lifetime performance of these strategies with $T_{u}$, comparison of  power efficiency and energy balancing in the network with update interval as $60$ and $3 \times 10^4$ packets are presented in Fig.\ref{fig3a} and Fig.\ref{fig3b},  respectively. Fig.\ref{fig3a} shows the comparison of energy consumption per packet for different strategies with $T_{u}$. In Fig.\ref{fig3b}, network wasted energy, which is the total remaining energy in the network after network expires, has been compared for different strategies with $T_{u}$. While strategies based on GLM policy require higher transmit power compared to MWTP MWM, MWTP SRS, network wasted energy for GLM policy based strategies is significantly low and therefore these strategies achieve higher network lifetime performance than MWTP MWM, MWTP SRS. The transmit energy consumption of GLM BM is high at low $T_{u}$ which accounts for its lower lifetime performance compared to GLM MBM, GLM SRS at low $T_{u}$.

\section{CONCLUSION}
In this paper, we have investigated the relay assignment and power allocation problem for GLM and MWTP policies to maximize lifetime of the SER constrained multiple-user cooperative networks. Optimal power allocation for source relay pair for each policy has been derived. The optimal relay assignment for MWTP and GLM policies have been solved using MWM algorithm with time complexity $\mathcal{O}(N^{3} )$ and BM algorithm with time complexity  $\mathcal{O}(N^{2.5} )$, respectively. Since relay assignment with BM algorithm is not power efficient, we  propose MBM algorithm which solves relay assignment optimally in time $\mathcal{O}(MN^{2.5})$ for GLM policy. To reduce the complexity of relay assignment further, a suboptimal relay selection algorithm has been proposed with time complexity $\mathcal{O}(MN )$. Simulation results demonstrate that the strategies based on GLM policy, achieve better network lifetime compared to the strategies based on MWTP policy. The network lifetime has been shown to improve if the relay selection and power allocation of the proposed strategies are updated frequently. The performance of GLM BM strategy is worse as compared to GLM SRS, GLM MBM strategies at low update interval. GLM MBM strategy has the best network lifetime performance and GLM SRS strategy performs close to it.

\label{sec_conclusion}

\bibliographystyle{IEEEtran}
\bibliography{ref_lifetime_conf_grouping}

\end{document}